\documentclass[12pt]{article}
\usepackage{amssymb,amsmath,latexsym}
\title{Spectral averaging techniques
\\ for Jacobi matrices with matrix entries}

\author{Christian Sadel, Hermann Schulz-Baldes
\\
\\
{\small Department Mathematik, Universit\"at Erlangen-N\"urnberg, Germany}
}

\date{ }

\newtheorem{theo}{Theorem}

\newtheorem{proposi}{Proposition}
\newtheorem{lemma}{Lemma}
\newtheorem{coro}{Corollary}

\newcommand{\CC}{{\mathbb C}}
\newcommand{\NN}{{\mathbb N}}
\newcommand{\RR}{{\mathbb R}}

\newcommand{\LM}{{\mathbb L}}

\newcommand{\DM}{{\mathbb D}}

\newcommand{\UM}{{\mathbb U}}
\newcommand{\WM}{{\mathbb W}}

\newcommand{\Tr}{\mbox{\rm Tr}}
\newcommand{\Tt}{{\cal T}}

\newcommand{\Mm}{{\cal M}}
\newcommand{\Cc}{{\cal C}}
\newcommand{\Jj}{{\cal J}}

\newcommand{\one}{{\bf 1}}
\newcommand{\nul}{{\bf 0}}

\begin{document}

\maketitle

\begin{abstract}
A Jacobi matrix with matrix entries is a self-adjoint block tridiagonal matrix
with invertible blocks on the off-diagonals. Averaging over boundary conditions
leads to explicit formulas for the averaged spectral measure which can potentially be useful for spectral analysis. Furthermore another variant of spectral averaging over coupling constants for these operators is presented.
\end{abstract}

\vspace{.5cm}

\section{Introduction}

\vspace{.2cm}

Many variants of spectral averaging for one-dimensional Sturm-Liouville or Jacobi operators are known \cite{CL}. If such operators depend on some continuous parameters, then the spectral averaging principle states that the spectral measures averaged over these parameters with respect to a measure with density are themselves absolutely continuous. In refinements useful for a detailed spectral analysis, it is possible to prove that they are even equivalent to the Lebesgue measure \cite{dRT,dRMS}. The continuous parameters are typically boundary conditions or coupling constants.

\vspace{.2cm}

For Jacobi operators with matrix entries the only contribution seems to be due to
Carmona and Lacroix \cite{CL}. Unfortunately, their work does not give all the
details of proof and the presentation is not conceptually structured nor does
it cover full generality. Part of this work, in particular Theorem~\ref{theo-boundav},
is thought to fill these gaps. The main ingredient of the proof is the Cauchy
formula for integration over the unitary group as proven by Hua \cite{Hua} (it is recalled in an appendix).
Theorem~\ref{theo-boundav} leads to a formula (stated in Theorem~\ref{theo-Carmona})
establishing a close link between spectral properties of the Jacobi operators
in the limit point case and their formal solutions expressed in terms of the transfer matrices. We believe that Theorem~\ref{theo-Carmona} can potentially be a useful alternative tool (other than Kotani theory \cite{KS}) for proving existence of absolutely continuous spectrum.  Finally Theorem~\ref{theo-boundav2} provides a matrix version of a well-known identity of rank one perturbation theory, showing that averaging of the spectral measure over both boundary conditions leads to the Lebesgue measure. As an
application, spectral stability results w.r.t. local perturbations are
presented. It is shown that also averages over fewer parameters than the whole set of boundary conditions lead to averaged spectral measures which are equivalent to the Lebesgue measure, at least locally in energy. This last part generalize the results in \cite{dRMS}. As we lack a subordinacy theory for Jacobi matrices with matrix entries the applications to spectral theory of that paper do not carry over.

\vspace{.1cm}

This work complements our prior works \cite{SB}
on Sturm-Liouville oscillation theory and \cite{SB2} on Weyl
theory for Jacobi matrices with
matrix entries, so the basic notations and setup are chosen accordingly. Heavy
use is being made of the matrix M\"obius transformation on which there is an
abundant literature (see the references in \cite{SB,SB2}), but the main facts
relevant for the present purposes are resembled in an appendix and all their
short proofs are given in \cite{SB,SB2}.

\vspace{.3cm}

\noindent {\bf Acknowledgment:} This work was supported by the DFG. We also thank the Newton Institute for hospitality.

\section{Setup and review of needed results}

\vspace{.2cm}

\noindent {\bf Notations:} The matrix entries of the Jacobi matrices are of
size $L\in\NN$. Matrices of size $L\times L$ are denoted by roman letters,
those of size $2L\times 2L$ by calligraphic ones. The upper half-plane $\UM_L$
is the set of complex $L\times L$ matrices satisfying $\imath(Z^*-Z)>\nul$. Its
closure $\overline{\UM_L}$ is given by matrices satisfying
$\imath(Z^*-Z)\geq\nul$. The boundary is a stratified space
$\partial\UM_L=\cup_{l=1}^L\partial_l\UM_L$, where $\partial_l\UM_L$ contains
those matrices in $\overline{\UM_L}$ for which the kernel of $Z^*-Z$ is
$l$-dimensional.

\vspace{.2cm}

\subsection{The Jacobi matrix and its resolvent}
\label{sec-resol}

\vspace{.2cm}

Fix two integers $L,N\in \NN$ and let $(T_n)_{n=2,\ldots,N}$ and
$(V_n)_{n=1,\ldots,N}$ be sequences of respectively invertible and self-adjoint
$L\times L$ matrices with complex entries. Furthermore let the left and right
boundary conditions $\hat{Z}$ and $Z$ be also self-adjoint $L\times L$
matrices. Then the associated Jacobi matrix with matrix entries
$H^N_{\hat{Z},Z}$ is by definition the self-adjoint operator acting on states
$\phi=(\phi_n)_{n=1,\ldots,N}\in \ell^2(1,\ldots,N)\otimes \CC^L$ by
\begin{equation}
\label{eq-jacobi} (H^N_{\hat{Z},Z}\,\phi)_n \;=\;
T_{n+1}\phi_{n+1}\,+\,V_n\phi_n \,+\,T_{n}^*\phi_{n-1} \;, \qquad
n=1,\ldots,N\;,
\end{equation}
where $T_1=T_{N+1}={\bf 1}$, together with the boundary conditions
\begin{equation}
\label{eq-boundary} \phi_0\;=\;\hat{Z}\,\phi_1\;, \qquad
\phi_{N+1}\;=\;-\,Z\,\phi_N\;.
\end{equation}
If $\hat{Z}={\bf 0}$ and ${Z}={\bf 0}$, one speaks of Dirichlet boundary
conditions at the left an right boundary respectively. It will be useful to
allow also non-selfadjoint boundary conditions $\hat{Z},Z\in\overline{\UM_L}$
hence giving rise to a possibly non-selfadjoint operator $H^N_{\hat{Z},Z}$. One
can rewrite $H^N_{\hat{Z},Z}$ as an $NL\times NL$ matrix with $L\times L$ block
entries:
\begin{equation}
\label{eq-matrix} H^N_{\hat{Z},Z} \;=\; \left(
\begin{array}{ccccccc}
V_1-\hat{Z}       & T_2  &        &        &         &        \\
T_2^*      & V_2    &  T_3  &        &         &        \\
            & T_3^* & V_3    & \ddots &         &        \\
            &        & \ddots & \ddots & \ddots  &        \\
            &        &        & \ddots & V_{N-1} & T_N   \\
            &        &        &        & T_N^*  & V_N -Z
\end{array}
\right)
\;.
\end{equation}
At times, our interest will only be in the dependence of the right boundary
condition $Z$, and then the index $\hat{Z}$ will be suppressed.

\vspace{.2cm}

As for a one-dimensional Jacobi matrix, it is useful to rewrite the eigenvalue
equation
\begin{equation}
\label{eq-Schroedinger} (H^N_{\hat{Z},{Z}}\,\phi)_n \;=\; z\,\phi_n\;, \qquad
n=1,\ldots,N\;,
\end{equation}
for a complex energy $z\in \CC$
in terms of the $2L\times 2L$ transfer matrices $\Tt_n^z$ defined by
\begin{equation}
\label{eq-transfer}
\Tt_n^z
\;=\;
\left(
\begin{array}{cc}
(z\,{\bf 1}\,-\,V_n)\,T_n^{-1} & - T_n^* \\
T_n^{-1} & {\bf 0}
\end{array}
\right)
\;,
\qquad
n=1,\ldots,N
\;,
\end{equation}
namely
\begin{equation}
\label{eq-transfersol}
\left(
\begin{array}{c}
T_{n+1}\phi_{n+1} \\
\phi_n
\end{array}
\right)
\;=\;
\Tt^z_n\,
\left(
\begin{array}{c}
T_{n}\phi_{n} \\
\phi_{n-1}
\end{array}
\right)
\;,
\qquad
n=1,\ldots,N
\;.
\end{equation}
This gives a solution of the eigenvalue equation \eqref{eq-Schroedinger} which,
however, does not necessarily satisfy the boundary condition
\eqref{eq-boundary}. Now $z\in\CC$ is an eigenvalue of $H^N_{\hat{Z},Z}$ if and
only if there is a solution of \eqref{eq-Schroedinger}, that is produced by
\eqref{eq-transfersol}, which satisfies \eqref{eq-boundary}. As is
well-established, one can understand \eqref{eq-boundary} as requirement on the
solution at sites $0,1$ and $N,N+1$ respectively to lie in $L$-dimensional
planes in $\CC^{2L}$. The corresponding two planes are described by the two
$2L\times L$ matrices (one thinks of the $L$  columns as spanning the plane)
\begin{equation}
\label{eq-boundaryplanes} \hat{\Phi}_{\hat{Z}} \;=\; \left(
\begin{array}{c}
{\bf 1} \\
-\,\hat{Z}
\end{array}
\right)\;, \qquad \Phi_Z \;=\; \left(
\begin{array}{c}
-\,Z \\
{\bf 1}
\end{array}
\right)\;.
\end{equation}
Then the boundary conditions (\ref{eq-boundary}) can be rewritten as
\begin{equation}
\label{eq-boundary2}
\left(
\begin{array}{c}
T_1\phi_{1} \\
\phi_0
\end{array}
\right) \;\in\; \hat{\Phi}_{\hat{Z}} \,\CC^L \;, \qquad \left(
\begin{array}{c}
T_{N+1}\phi_{N+1} \\
\phi_N
\end{array}
\right) \;\in\; \Phi_Z \,\CC^L \;.
\end{equation}
One way to attack the eigenvalue problem is to consider the $L$-dimensional
plane $\hat{\Phi}_{\hat{Z}}$ as the initial condition for an evolution of
$L$-dimensional planes under the application of the transfer matrices:
\begin{equation}
\label{eq-Lagdyn} \Phi_n^z \;=\; \Tt_n^z\,\Phi_{n-1}^z \;, \qquad \Phi_0^z
\;=\; \hat{\Phi}_{\hat{Z}} \;.
\end{equation}
Because the transfer matrices are invertible, this produces an $L$-dimensional
set of solutions of \eqref{eq-transfersol}. With the correspondence
\begin{equation}
\label{eq-solcorr} \Phi^{z}_n\;=\; \left(
\begin{array}{c}
T_{n+1}\, \phi^{z}_{n+1}\\
\phi^{z}_n
\end{array}
\right)\;,
\end{equation}
this also gives a matricial solution $\phi_n^z$ of \eqref{eq-Schroedinger}. Due
to the initial condition in \eqref{eq-Lagdyn} the left boundary condition at
sites $0,1$ is automatically satisfied. The dimension of the intersection of
the plane $\Phi_N^z$ with the plane $\Phi_Z$ gives the number of linearly
independent solutions of \eqref{eq-Schroedinger} at energy $z$, and therefore
the multiplicity of $z$ as eigenvalue of $H^N_{\hat{Z},Z}$.

\vspace{.2cm}

Given \eqref{eq-Lagdyn}, but also its own sake, it is natural to introduce the
transfer matrices over several sites by
\begin{equation}
\label{eq-iterate} \Tt^z(n,m)\;=\; \Tt_n^z\cdot\ldots\cdot\Tt^z_{m+1}\;,\qquad
n>m \;,
\end{equation}
as well as $\Tt^z(n,n)={\bf 1}$ and $\Tt^z(n,m)=\Tt^z(m,n)^{-1}$ for $n<m$.
With this notation, the solution of the eigenvalue equation
\eqref{eq-Schroedinger} satisfies $\Phi^z_n=\Tt^z(n,m)\Phi_m^z$ and, in
particular, $\Phi^z_n=\Tt^z(n,0)\hat{\Phi}_{\hat{Z}}$.
Of particular importance will be the transfer matrix $\Tt^z(N,0)$
across the whole sample. Let us introduce the notations
\begin{equation}
\label{eq-entrydef} \Tt^z(N,0) \;=\; \left(
\begin{array}{cc} A^z_N & B^z_N \\ C^z_N & D^z_N
\end{array}
\right) \;,
\end{equation}
where all entries are $L\times L$ matrices. These matrices will intervene in
many of the results below. Let us point out that $\Tt^z(N,0)$ and all its
entries do not depend on the boundary conditions $\hat{Z}$ and $Z$. The transfer matrix including boundary conditions is then
\begin{equation}
\label{eq-includeBC}
\left(
\begin{array}{cc}
{\bf 1} & Z \\
\nul & \one
\end{array}
\right) \Tt^z(N,0) \left(
\begin{array}{cc}
{\bf 1} & \nul \\
-\,\hat{Z} & \one
\end{array}
\right)
\;=\; \left(
\begin{array}{cc} A^z_N + ZC^z_N -B^z_N \hat{Z}- ZD^z_N\hat{Z}
& B^z_N + ZD^z_N \\ C^z_N - D^z_N\hat{Z} & D^z_N
\end{array}
\right)
\:.
\end{equation}

\vspace{.2cm}

Now we introduce the resolvent. Let
$\pi_n:\CC^L\to \CC^{NL}$ for $n=1,\ldots,N$ denote the partial isometry
$$
\pi_n|l\rangle\;=\;|n,l\rangle\;,\qquad l=1,\ldots,L\;,
$$
where the Dirac notation for localized states in $\CC^{N}\otimes\CC^{L}$ is
used. Then the $L\times L$ Green's matrix is given by
$$
G^z_N(\hat{Z},Z,n,m) \;=\;\pi_n^* (H^N_{\hat{Z},Z}-z\,\one)^{-1}\pi_m\;.
$$

\vspace{.2cm}

\begin{proposi}
\label{prop-Greenformulas2} {\rm \cite{SB2}} For
$\hat{Z},Z\in\overline{\UM_L}$,
\begin{eqnarray*}
G^z_N(\hat{Z},Z,1,1) & = &
\left[A^z_N+Z C^z_N -B^z_N\hat{Z} -Z D^z_N\hat{Z}\right]^{-1}
\left[B^z_N +Z D^z_N\right]  \nonumber
\\
& = &
\left[\left[A^{\overline{z}}_N+Z C^{\overline{z}}_N -B^{\overline{z}}_N\hat{Z} -Z D^{\overline{z}}_N\hat{Z}\right]^{-1}
\left[B^{\overline{z}}_N +Z D^{\overline{z}}_N\right]  \right]^*
\;.\nonumber
\end{eqnarray*}
\end{proposi}

\vspace{.2cm}

\subsection{Parametrization of the boundary conditions}
\label{sec-symplectic}

\vspace{.2cm}

The underlying hermitian symplectic structure is an important ingredient in
most of the equations of the last sections, in particular in their proofs. It
is necessary in order to understand what the adequate spectral averaging over
the boundary conditions is. This section first recalls basic fact about the
symplectic structure, which will then be applied below. Let the symplectic form
$\Jj$ be the $2L\times 2L$ matrix defined by
$$
\Jj\;=\;
\left(
\begin{array}{cc}
0 & -\one \\ \one & 0
\end{array}
\right)\:.
$$
An $L$-dimensional
plane described by a $2L\times L$ matrix $\Phi$ of maximal rank is called
Lagrangian (or also isotropic, or simply symplectic) if $\Phi^*\Jj\Phi={\bf
0}$.

\vspace{.2cm}

Two $L$-dimensional planes described by $2L\times L$ matrices $\Phi$ and $\Psi$
are called equivalent if there exists $c\in\,$Gl$(L,\CC)$ with $\Phi=\Psi c$.
The Lagrangian Grassmannian $\LM_L$ is by definition the set of equivalence
classes of Lagrangian planes. It is difficult to track back the original
reference for the following result (it probably predates \cite{Bot}). A short
proof can be found in \cite{SB} where it is also shown how two natural
symmetries are implemented.

\vspace{.2cm}

\begin{proposi}
\label{prop-parametrize} The Lagrangian Grassmannian $\LM_L$ is identified
with the unitary group {\rm U}$(L)$ via the real analytic diffeomorphism
$\Pi:\LM_L\to\,${\rm U}$(L)$ given by
$$
\Pi([\Phi]_\sim)
\;=\;
(a-\imath b)(a+\imath b)^{-1}\;,
\qquad
\Phi
\;=\;\left(
\begin{array}{c}
a \\ b
\end{array}
\right)
\;.
$$
\end{proposi}

\vspace{.2cm}

Due to this theorem there is a natural measure on the Lagrangian Grassmannian
$\LM^\CC_L$ given by the pull-back under $\Pi$ of the Haar measure on the
unitary group.

\vspace{.2cm}

The Lie group conserving the (hermitian) symplectic structure is the
(hermitian) symplectic group SP$(2L,\CC)$ defined by those $2L\times 2L$
matrices $\Tt$ satisfying $\Tt^*\Jj\Tt=\Jj$.
Clearly, if $\Phi$ describes a Lagrangian plane, then so does $\Tt\Phi$ for any
$\Tt\in\,$SP$(2L,\CC)$. Isomorphic to the hermitian symplectic group is the
Lorentz group of signature $(L,L)$
defined by U$(L,L,\CC)=\Cc\,$SP$(2L,\CC)\Cc^*$, where $\Cc$ is
the Cayley transformation introduced as the matrix
$$
\Cc
\;=\;\frac{1}{\sqrt{2}}\;
\left(
\begin{array}{cc} {\bf 1} & -\,\imath\,{\bf 1} \\
{\bf 1} & \imath\,{\bf 1} \end{array}
\right)
\;.
$$

\vspace{.2cm}

Next let us exhibit explicitly the symplectic structure in the equations of
Section~\ref{sec-resol}. Both of the planes $\hat{\Phi}_{\hat{Z}}$ and
$\Phi_Z$ used as boundary conditions in \eqref{eq-boundary2} are Lagrangian
in the above sense.  Actually there are many Lagrangian
planes which cannot be written in this way, but they form a set of zero
measure. Due to Proposition~\ref{prop-parametrize}, it is natural to identify the
left and right boundary conditions with unitary matrices:
\begin{equation}
\label{eq-boundpar} \hat{U}\;=\;\Pi([\hat{\Phi}_{\hat{Z}}]_\sim)\;, \qquad
{U}\;=\;\Pi([{\Phi}_{{Z}}]_\sim)\;.
\end{equation}
In other terms, this means $U=\Cc\cdot(-Z)$ and $\hat{U}=-\Cc\cdot \hat{Z}$.
Furthermore, let us set
$$
U^E_n\;=\;\Pi([\Phi_n^E]_\sim)\;,
$$
where $\Phi_n^E$ is the solution \eqref{eq-Lagdyn} which automatically
satisfies the left boundary condition. Then $U^E_0=\hat{U}$. Using the correspondence~\eqref{eq-boundpar}, we also set
$$
G^z_N(\hat{U},U)\;=\;G^z_N(\hat{Z},Z,1,1)\:.
$$
As $z\mapsto G^z_N(\hat{U},U)\in\UM_L $ is analytic in $z$ for
$\Im m(z)>0$, the Herglotz representation theorem \cite{GT} associates a
matrix-valued (averaged spectral) measure:
$$
G^z_N(\hat{U},U)
\;=\;\int \rho^N_{\hat{U},U}(dE)\;\frac{1}{E-z}\;.
$$

\vspace{.2cm}

\subsection{The oscillation theorem}

\vspace{.2cm}

The oscillation theorem is another application of the parametrization of
boundary conditions. It is stated for sake of completeness and because it will
be used in the proof of the result of Section~\ref{sec-aver}. Due to
Section~\ref{sec-symplectic}, $E\in\RR\mapsto \Phi_N^E$ is a path of Lagrangian
planes and for each $E$ the dimension of its intersection with the right
boundary condition $\Phi_Z$ is the multiplicity of $E$ as an eigenvalue of
$H^N_{\hat{Z},Z}$. This intersection number was introduced by Bott
\cite{Bot} precisely for the study of the eigenvalue calculation of
Sturm-Liouville operators, the continuous analogues of Jacobi matrices. Later
on it was rediscovered by Maslov and a detailed survey of its properties is
included in \cite{SB}. The intersection number can be conveniently calculated
using the associated unitary $U^E_N$ and this leads to the following theorem
which was proven in \cite{SB} under the supplementary hypothesis that the
$T_n$'s are positive, but the proof directly transposes to the slightly
generalized situation considered here.

\vspace{.2cm}

\begin{theo}
\label{theo-osci} Let $E\in\RR$, $N\geq 2$, and {\rm (}for sake of
simplicity{\rm )} the right boundary condition be Dirichlet, that is
$Z=\nul$. Then there are $L$ strictly increasing real analytic functions
$\theta^E_{N,l}:\RR\to\RR$, $l=1,\ldots,L$, such that $e^{\imath
\theta^E_{N,l}}$ are the eigenvalues of $U_N^E$. The multiplicity of $E$ as an
eigenvalue of $H^N_{\hat{Z},\nul}$ is equal to the multiplicity of $-1$ as an
eigenvalue of $U_N^E$. Furthermore, the matrix
$\frac{1}{\imath}\,(U^E_N)^*\partial_EU^E_N$ is positive definite.
\end{theo}

\vspace{.2cm}

\subsection{Limit point operators}

\vspace{.2cm}

If in the prior sections $N=\infty$, then the right boundary condition $Z$ is pushed to infinity. If this gives a well-defined (essentially self-adjoint) operator $H_{\hat{Z}}$, one speaks of the limit point case. Various criteria for this can be given, the simplest one being that $\|T_n\|$ is uniformly bounded from below. Otherwise one needs the infinite operator having non-vanishing deficiency spaces and one has to consider various self-adjoint extensions. Here we restrict ourself to limit point operators. For these operators, limits
$$
G^z(\hat{U})\;=\;\lim_{N\to\infty}\;G^z_N(\hat{U},U)\;,
$$
exist, are independent of $U$ and are the Green function of $H_{\hat{Z}}$. Its spectral measure is denoted by $\rho_{\hat{U}}$ and obtained as the weak limit of $\rho^N_{\hat{U},U}$.

\vspace{.2cm}

\section{Average over boundary conditions}

\vspace{.2cm}

Let us write $dU$ for the normalized Haar measure on
U$(L)$.

\begin{theo}
\label{theo-boundav}
For $\Im m(z)>0$, one has
\begin{eqnarray}
\label{eq-boundav1}
\int dU\;G^z_N(\one,U)
& = &
\left[
A^z_N+\imath\, C^z_N
\right]^{-1}
\left[B^z_N+\imath\, D^z_N
\right]\; \\
& = & \left[(B^{\overline{z}}_N)^*+\imath(D^{\overline{z}}_N)^*\right]\,
\left[(A^{\overline{z}}_N)^*+\imath (C^{\overline{z}}_N)^*\right]^{-1} \;.
\label{eq-boundav11}
\end{eqnarray}
Moreover, for $E=\Re e(z)$, and setting $\Im m(A)=\frac{1}{2\imath}(A-A^*)$ for any square matrix $A$,
\begin{equation}
\label{eq-boundav2}
\lim_{\Im m(z)\downarrow 0}\;\int dU\;\Im m(
G^z_N(\one,U))\;=\;
\left[
(A^E_N)^*A^E_N+(C^E_N)^* C^E_N
\right]^{-1}
\;.
\end{equation}
\end{theo}

\vspace{.2cm}

\noindent {\bf Remark 1} A formula similar to \eqref{eq-boundav2} can be found
in \cite{CL}, but the latter authors use the average over the Haar measure on
the symmetric space of symmetric unitaries instead of the group U$(L)$
(moreover, their proof seems to have several gaps).

\vspace{.1cm}

\noindent {\bf Remark 2} It is easy to incorporate
the left boundary condition $\hat{U}\not = \one$ using \eqref{eq-includeBC}.

\vspace{.1cm}

\noindent {\bf Remark 3} One way to define the closed
Weyl disc $\overline{\WM^z_N}$ is as the image of the map
$Z\in \overline{\UM_L} \mapsto  G_N^z(\one,Z,1,1)$. As proven in \cite{SB2}
the points in the Weyl disc can also be parametrized by
$G_N^z(\one,Z,1,1)=S^z_N +(R^z_N)^{\frac{1}{2}}W
(R^{\overline{z}}_N)^{\frac{1}{2}}$ where $S^z_N$ and $R^z_N>0$ are
properly defined center and radial operators, and $W\in\,$U$(L)$ depends on $Z$, {\it cf.} \cite{SB2}.
Taking the average over $W$ w.r.t. to the Haar measure in this
representation immediately gives $\int dW\,G_N^z(\one,Z,1,1)=S^z_N$,
which is {\it not}
equal to the r.h.s. of \eqref{eq-boundav1}. The Jacobian of the change of
variables $Z\mapsto W$ does not seem to be known (nor be of great importance).

\vspace{.2cm}

\noindent {\bf Proof} of Theorem~\ref{theo-boundav}. First let us note that one
can use the M\"obius transformation to express
$U=\Pi([\Phi_Z]_\sim)=\Cc\cdot(-\,Z)$. Hence also $Z=-\,\Cc^*\cdot U$.
Starting from Proposition~\ref{prop-Greenformulas2}, one therefore has
$$
\int dU\;G^z_N(\one,U) \;=\;
\int dU\;\left( A^z_N- \Cc^*\cdot U\;C^z_N \right)^{-1}
\left(B^z_N-\Cc^*\cdot U\; D^z_N \right)\;.
$$
By \eqref{eq-reflext},
$$
\int dU\;G^z_N(\one,U) \;=\; \int dU\;\left( A^z_N+\, \Cc^*\cdot U\;C^z_N
\right)^{-1} \left(B^z_N+\Cc^*\cdot U\; D^z_N \right)\;.
$$
(Alternatively to this argument, one could have defined the average on the
l.h.s. by the r.h.s..) In order to be able to apply the Cauchy formula
\eqref{eq-Cauchy} for $Z=\nul$, it is sufficient to show the analyticity of the
function
$$
f(Z)\;=\; ( A^z_N+ \Cc^*\cdot Z\,C^z_N )^{-1} (B^z_N+\Cc^*\cdot Z\, D^z_N )\;,
$$
on the unit disc ${\DM^\CC_L}$ as well as its continuity on the closure
$\overline{\DM^\CC_L}$ (strictly speaking, one should consider the entries of
the matrix-valued function $f$). This follows from Weyl theory \cite{SB2}
combined with the fact that
$(-\,\Cc^*\cdot Z)$ is in the closed lower half plane (for
$Z\in\overline{\DM^\CC_L}$). The Cauchy formula \eqref{eq-Cauchy} for $Z=\nul$
now concludes the proof of \eqref{eq-boundav1} because $\Cc^*\cdot \nul =
\imath\,\one$. Formula \eqref{eq-boundav11} is proven similarly from the second identity in Proposition~\ref{prop-Greenformulas2}.

\vspace{.1cm}

It follows from the results of \cite{SB2} that
$A_N^z+\imath\,C_N^z=(A_N^z(C_N^z)^{-1}+\imath\,\one)C_N^z$ is invertible.
Inserting $\one=[(A_N^z+\imath\,C_N^z)^*]^{-1}(A_N^z+\imath\,C_N^z)^*$ in
\eqref{eq-boundav1} shows
\begin{eqnarray*}
\int dU\;G^z_N(\one,U)  & = & \left[ (A^z_N)^*A^z_N+(C^z_N)^* C^z_N +\imath\,
((A^z_N)^*C^z_N-(C^z_N)^*A^z_N)\right]^{-1}\\
& & \;\; \left[(A^z_N)^*B^z_N+(C^z_N)^* D^z_N + \imath
\,((A^z_N)^*D^z_N-(C^z_N)^*B^z_N) \right]\;.
\end{eqnarray*}
As the transfer matrices at real energies are symplectic, the limit of
vanishing imaginary part in the energy can be taken in this equation and that
directly implies
$$
\lim_{\Im m(z)\downarrow 0}\;
\int dU\;G^z_N(\one,U)  \; = \;
\left[ (A^E_N)^*A^E_N+(C^E_N)^* C^E_N \right]^{-1}
\;\left[(A^E_N)^*B^E_N+(C^E_N)^* D^E_N + \imath
\,\one \right]\;,
$$
where we used the identity $(A^E_N)^*D^E_N+(C^E_N)^* B^E_N=\one$ holding for any symplectic matrix.
The same calculation can be carried out starting from
\eqref{eq-boundav11} and adding the results up gives
\eqref{eq-boundav2}. \hfill $\Box$

\vspace{.2cm}

One corollary of Theorem~\ref{theo-boundav} is the following formula which
links the averaged spectral measure defined by
\begin{equation}\label{eq-avmeas1}
\rho^N_{\hat{U}}\;=\;\int dU\,\rho^N_{\hat{U},U}
\end{equation}
to properties of the eigenfunctions of the transfer
matrices at real energies.

\begin{coro}
\label{coro-solspec0} For any $E_0<E_1$,
%
$$
\frac{1}{2}\,\bigl[\,\rho^N_{\hat{U}}([E_0,E_1])+
\rho^N_{\hat{U}}((E_0,E_1))\, \bigr]
\;=\;\int^{E_1}_{E_0}dE\; \left[\,|A^E_N-B^E_N\Cc^*\!\cdot(-\hat{U})|^2 +
|C^E_N-D^E_N\Cc^*\!\cdot(-\hat{U})|^2\,\right]^{-1}\,.
$$
%
\end{coro}

In the limit point case, the averaging in \eqref{eq-avmeas1} becomes irrelevant because
$\rho^N_{\hat{U},U}$ converges weakly to the spectral measure $\rho_{\hat{U}}$ of
$H_{\hat{Z}}$ as $N\to\infty$. This leads to the following formula for the spectral measure, which was already obtained by Carmona in the strictly one-dimensional case $L=1$, and by Pearson \cite{Pea} for one-dimensional Schr\"odinger operators.
For sake of simplicity, let us set $\hat{U}=\one$ so that $\Cc^*\!\cdot(-\hat{U})=0$.

\begin{theo}
\label{theo-Carmona} Let the semi-infinite Jacobi matrix $H$ be in the limit point case.
Then, for any $E_0<E_1$,
\begin{equation}\label{eq-solspec}
\frac{1}{2}\,\bigl[\,\rho_{\one}([E_0,E_1])+\rho_{\one}((E_0,E_1))\, \bigr]
\;=\;\lim_{N\to\infty} \;\int^{E_1}_{E_0}dE\; \left[\,(A^E_N)^*A^E_N+(C^E_N)^*
C^E_N\,\right]^{-1}\;.
\end{equation}
\end{theo}

Next we also average over the left boundary condition $\hat{U}$. The associated averaged spectral measure is equal to the Lebesgue measure, a fact also known from the case $L=1$.

\begin{theo}
\label{theo-boundav2}
For any $N$ and $H^N$,
\begin{equation}
\label{eq-boundav3}
4\;\int d\hat{U}\;(\hat{U}-\one)^{-1}\;
\rho^N_{\hat{U}}(dE)\;
(\hat{U}^*-\one)^{-1}
\;=\;\one\;dE\;.
\end{equation}
\end{theo}

\vspace{.2cm}

The proof of the theorem is based on the following integral identity.

\begin{lemma}
\label{lem-radav} Let $0<\Tt\in\mbox{\rm U}(L,L)$ and $V\in \mbox{\rm U}(L)$.
Then
$$
\int dU\;
\left[\;\left(
\begin{array}{c} U \\  V \end{array}
\right)^*
\Tt
\left(
\begin{array}{c} U \\  V \end{array}
\right)
\;\right]^{-1}\;=\;\one\;.
$$
\end{lemma}

\noindent {\bf Proof.} Let $I$ denote the integral appearing in the lemma.
First let us use that $0<\Tt\in\mbox{\rm U}(L,L)$
can be transformed into a normal form by $\Mm=\left(
\begin{array}{cc} W & 0 \\  0 & W' \end{array}
\right)\in \mbox{\rm U}(L,L)\cap
\mbox{\rm U}(2L)\cong\mbox{\rm U}(L)\oplus\mbox{\rm U}(L)$ where $W,W'\in \mbox{\rm U}(L)$, namely
$$
\Mm^*\Tt\Mm
\;=\;
\left(
\begin{array}{cc} \cosh(\eta) & \sinh(\eta)
\\ \sinh(\eta)  & \cosh(\eta) \end{array}
\right)\;.
$$
where $\eta={\rm diag}(\eta_1,\ldots,\eta_L)$ is a diagonal matrix with non-negative entries.
Let us denote the r.h.s. by $\Tt_\eta$.
Replacing this identity, one obtains
$$
I\;=\:
(W'V)^*\int dU\;
\left[\;\left(
\begin{array}{c} WU(W'V)^* \\  \one \end{array}
\right)^*
\Tt_\eta
\left(
\begin{array}{c} WU(W'V)^* \\  \one \end{array}
\right)
\;\right]^{-1}(W'V)\:,
$$
so that using the invariance of the Haar measure $dU$ one realizes that it is sufficient to consider the case $V=\one$ and $\Tt=\Tt_\eta$.
First suppose $\eta>0$. In this case, $\sinh(\eta)$ is invertible and
\begin{eqnarray*}
I
&=&
\int dU \left[\cosh(\eta) U +  U\cosh(\eta)+U\sinh(\eta)U+\sinh\eta \right]^{-1}U\\
&=&
\int dU \left[\left(U+\frac{\cosh(\eta)+\one}{\sinh(\eta)}\right)^{-1}
\left(\sinh(\eta)\right)^{-1}
\left(U+\frac{\cosh(\eta)-\one}{\sinh(\eta)}\right)^{-1}U \right]\;.
\end{eqnarray*}
To simplify notations, define $\alpha=\frac{\cosh(\eta)+\one}{\sinh(\eta)}$
and $\beta=\frac{\cosh(\eta)-\one}{\sinh(\eta)}$. Note that
$0<\beta<\one<\alpha$, $\alpha\beta=\one$ and $(\alpha-\beta)\sinh(\eta)=2$.
Now using
$$
\int dU (U+\beta)^{-1} U
\;=\;
\int dU \sum_{n\geq 0} \left(-U^{-1}\beta\right)^n
\;=\;
\one\;,\qquad\int dU (U+\alpha)^{-1} U
\;=\;0\;,
$$
one gets with the resolvent identity
$$
I-\one
\; = \;
2
\int dU\left[
\left((U+\alpha)^{-1}-(\alpha-\beta)^{-1}\right)\sinh(\eta)^{-1} (U+\beta)^{-1}U
\right]
\;=\;-\int dU\,(U+\alpha)^{-1}U\;=\;0
\:,
$$
and the proof is complete in the case $\eta>0$. By continuity of the integral one also recovers the case $\eta\geq 0$.
\hfill $\Box$

\vspace{.2cm}

\noindent {\bf Proof} of Theorem~\ref{theo-boundav2}.
This is based on Corollary~\ref{coro-solspec0}. Actually it is sufficient to show that the average of the integrand on the r.h.s. of  Corollary~\ref{coro-solspec0} satisfies
$$
4\;\int d\hat{U}\;(\hat{U}-\one)^{-1}
\;\left[\,|A^E_N-B^E_N\Cc^*\!\cdot(-\hat{U})|^2 +
|C^E_N-D^E_N\Cc^*\!\cdot(-\hat{U})|^2\,\right]^{-1}
(\hat{U}^*-\one)^{-1}
\;=\;\one\;.
$$
(Note that the various inverse appearing in this formula do not exist on a set of zero measure.)
This identity reduces to
$$
2\;\int d\hat{U}\;\left[\,
\left(
\begin{array}{c} U \\  \one \end{array}
\right)^*\Cc \,\Tt^E(N,0)^*\Tt^E(N,0)\,\Cc^*
\left(\begin{array}{c} U \\  \one \end{array}
\right)\,\right]^{-1}
\;=\;\one\;.
$$
Now $\Tt^E(N,0)^*\Tt^E(N,0)$ is a positive symplectic matrix and hence
$\Cc \Tt^E(N,0)^*\Tt^E(N,0)\Cc^* $ is a positive matrix in U$(L,L)$. Hence
Lemma~\ref{lem-radav} shows that this identity indeed holds.
\hfill $\Box$

\section{Spectral measures averaged over coupling constants}
\label{sec-aver}

As an application of the results of Theorem~\ref{theo-Carmona} we consider here a
particular one-parameter family of Jacobi matrices with matrix entries obtained
by a local positive perturbation and show that the associated averaged spectral
measure is under certain conditions absolutely continuous.

\vspace{.2cm}

Let $H$ be a Jacobi matrix with matrix entries in the limit point case,
and let $\pi_n:\CC^L\to(\CC^L)^{\NN}$ denote the partial isometry onto the
$n$th site. For real positive semi-definite matrices $(W_n)_{1\leq n\leq N}$
and $\mu\in\RR$ define
$$
H(\mu)\;=\;H\,+\,\mu\;\sum_{n=1}^N\,\pi_n\,W_n\,\pi_n^*\;.
$$
Furthermore, let $H^N(\mu)$ be the finite Jacobi matrix obtained by projecting
$H(\mu)$ to the first $N$ sites. It is of the form \eqref{eq-matrix} with $V_n$
replaced by $V_n+\mu W_n$ and $\hat{Z}=Z=0$. Because the perturbation
$H(\mu)-H$ is increasing in $\mu$, the eigenvalues of $H(\mu)$ are increasing
functions of $\mu$. Finally let $\rho(\mu)$ be the matrix-valued spectral
measure of $H(\mu)$ and define the averaged
spectral measure corresponding to an interval $I=[\mu_0,\mu_1]$ by
$$
\overline{\rho}\;=\;\int_I d\mu\;\Tr(\rho(\mu))\;.
$$
%

\begin{theo}
\label{theo-specav} Suppose that $W_n>0$ and $W_{n+1}>0$ for some
$n=1,\ldots,N-1$. Let $I=[\mu_0,\mu_1]$ be sufficiently large such that there
are $2L$ eigenvalues of $H^N(\mu)$ passing by $E$ as $\mu$ varies in $I$. Then
$\overline{\rho}$ is equivalent to the Lebesgue measure in a neighborhood of
$E$.
\end{theo}

As can be seen from the proof below the hypothesis can be
somewhat relaxed. For $L=1$ the result was proven in \cite{dRMS}. Similar as in
\cite{dRMS}, the condition on the size of $I$ can also expressed in terms of an
associated Birman-Schwinger operator and, furthermore, it is also possible to
consider several parameter spectral averaging instead of over just one
parameter $\mu$. On the other hand, the applications to spectral analysis do
not carry over immediately, because the subordinacy theory is not yet developed
for Jacobi matrices with matrix entries.

\vspace{.2cm}

First we need to fix some notations. Just as $H(\mu)$, all objects of the
previous sections depend on a supplementary parameter $\mu$. In particular, we
will write $\Tt^z_n(\mu)$ and $\Tt^z(n,m,\mu)$. Furthermore, let us introduce the Dirichlet solutions ${\Psi}_N^{\mbox{\rm\tiny D},z}(\mu)=\Tt^z(N,0,\mu)\binom{\one}{0}$ and the matrix
$$
P^E_N(\mu)\;=\;-\,{\Psi}_N^{\mbox{\rm\tiny D},E}(\mu)^*\,\Jj\partial_\mu
{\Psi}_N^{\mbox{\rm\tiny D},E}(\mu)\;.
$$
The proof of Theorem~\ref{theo-specav} will be based on the following
criterion.

\begin{lemma}
\label{lem-criteria} Suppose:

\vspace{.1cm}

\noindent {\rm (i)} There exist positive constants $C_1,C_2$ such that
$C_1\,\one\leq P^E_N(\mu)\leq C_2\,\one$ for all $\mu\in I$.

\vspace{.1cm}

\noindent {\rm (ii)}
$$
\int_I\frac{d\mu}{2\pi}\;\Im m\;\partial_\mu\;\log\bigl( \Pi(
{\Psi}_N^{\mbox{\rm\tiny D},E}(\mu))\bigr)\;<\;-\,L\;.
$$

\vspace{.1cm}

\noindent Then $\overline{\rho}$ is equivalent to the Lebesgue measure in a
neighborhood of $E$.
\end{lemma}

\noindent {\bf Proof.} We write $\overline{\rho}(E_0,E_1)$ for
$\frac{1}{2}\,\bigl[\,\overline{\rho}([E_0,E_1])+\overline{\rho}((E_0,E_1))\,
\bigr]$. Let us start by integrating \eqref{eq-solspec} over $\mu$ and using
the dominated convergence theorem as well as Fubini's theorem:
$$
\overline{\rho}(E_0,E_1) \;=\;\lim_{M\to\infty} \;\int^{E_1}_{E_0}dE\;\int_I
d\mu \;\mbox{Tr}\left( \left|\, \Tt^E(M,N){\Psi}_N^{\mbox{\rm\tiny D},E}(\mu)
\,\right|^{-2}\right)\;.
$$
Now  for positive semi-definite operators $A,B$ with $0<C_1\,\one\leq B\leq
C_2\,\one$,
$$
\frac{1}{C_2}\;\mbox{Tr}(AB)\;\leq\; \mbox{Tr}(A)\;\leq\;
\frac{1}{C_1}\;\mbox{Tr}(AB)\;.
$$
Applying these bounds for $B=P^E_N(\mu)$ shows that
$$
\overline{\rho}(E_0,E_1) \;\approx\;\lim_{M\to\infty}
\;\int^{E_1}_{E_0}dE\;\int_I d\mu \;\mbox{Tr}\left( \left|\,
\Tt^E(M,N){\Psi}_N^{\mbox{\rm\tiny D},E}(\mu)
\,\right|^{-2}P^E_N(\mu)\,\right)\;,
$$
where the sign $\approx$ means that we have two-sided bounds. As
$\Tt^E(M,N)^*\Jj\Tt^E(M,N)=\Jj$, ${\Psi}_M^{\mbox{\rm\tiny
D},E}(\mu)=\Tt^E(M,N){\Psi}_N^{\mbox{\rm\tiny D},E}(\mu)$ and $\Tt^E(M,N)$ does
not depend on $\mu$, this can be rewritten as
\begin{eqnarray*}
\overline{\rho}(E_0,E_1) & \approx & -\,\lim_{M\to\infty}
\;\int^{E_1}_{E_0}dE\;\int_I d\mu \;\mbox{Tr}\left( \left|\,
{\Psi}_M^{\mbox{\rm\tiny D},E}(\mu) \,\right|^{-2} {\Psi}_M^{\mbox{\rm\tiny
D},E}(\mu)^*\,\Jj\partial_\mu {\Psi}_M^{\mbox{\rm\tiny D},E}(\mu) \,\right)\\
& = & -\,\pi\;\lim_{M\to\infty} \;\int^{E_1}_{E_0}dE\;\int_I \frac{d\mu}{2\pi}
\; \Im m\;\partial_\mu\;\log\bigl( \Pi( {\Psi}_M^{\mbox{\rm\tiny
D},E}(\mu))\bigr) \;,
\end{eqnarray*}
where the second identity is checked in \cite[Lemma 4]{SB}. Now the expression
under the integral $\int dE$ on the r.h.s. is precisely the pairing
$\int_\Gamma \omega$ of the Arnold cocycle $\omega$ with the path
$\Gamma(\mu)={\Psi}_M^{\mbox{\rm\tiny
D},E}(\mu)=\Tt^E(M,N){\Psi}_N^{\mbox{\rm\tiny D},E}(\mu)$, $\mu\in I$, in the
Lagrangian Grassmannian $\LM_L$ (actually here this is a path in the real Lagrangian Grassmannian
because $H(\mu)$ is real). Hypothesis states something about the pairing with the path
$\Gamma'(\mu)={\Psi}_N^{\mbox{\rm\tiny D},E}(\mu)$, namely
$\int_{\Gamma'}\omega <-L$. However, these two paths are related by the
multiplication with the symplectic matrix $\Tt^E(M,N)$. Hence by
\cite[Proposition 4]{SB}:
$$
\left|\;\int_{\Gamma}\omega\;-\;\int_{\Gamma'}\omega\;\right|\;\leq\;L\;.
$$
Therefore $0<C_3<-\int_{\Gamma}\omega <C_4$ where the upper bound follows from
compactness of $I$ and the constants are independent of $M$. Replacing this shows
$$
\overline{\rho}(E_0,E_1) \; \approx \; \lim_{M\to\infty} \;\int^{E_1}_{E_0}dE
\;=\;E_1-E_0\;,
$$
which is precisely the claimed equivalence of $\overline{\rho}$ with the
Lebesgue measure.
\hfill $\Box$

\vspace{.2cm}

\noindent {\bf Proof} of Theorem~\ref{theo-specav}. First of all,
$$
\Tt^E(N,0,\mu)^*\Jj^*\partial_\mu\Tt^E(N,0,\mu)
\;=\;\sum_{n=1}^N\;\Tt^E(n-1,0,\mu)^*\,\Tt^E_n(\mu)^*\Jj^*
(\partial_\mu\Tt^E_n(\mu))\,\Tt^E(n-1,0,\mu)\;.
$$
But
$$
\Tt^E_n(\mu)^*\Jj^* \partial_\mu\Tt^E_n(\mu)\;=\; \left(\begin{array}{cc}
(T_n^{-1})^*W_n\,T_n^{-1} & 0 \\ 0 & 0
\end{array}\right)\;
$$
is positive semi-definite, and the arguments in the proof of \cite[Proposition
6]{SB} show that
$$
\left(\begin{array}{cc} (T_{n+1}^{-1})^*W_{n+1}\,T_{n+1}^{-1} & 0 \\ 0 & 0
\end{array}\right)
\; + \;\Tt^E_{n+1}(\mu)^*\,\left(\begin{array}{cc} (T_n^{-1})^*W_n\,T_n^{-1} &
0
\\ 0 & 0
\end{array}\right)
\,\Tt^E_{n+1}(\mu)\;>\;0
$$
(strict positivity), whenever $W_n>0$ and $W_{n+1}>0$. In the latter case the
above sum is therefore strictly positive. Hence the hypothesis of
Theorem~\ref{theo-specav} imply $P^E_N(\mu)\geq C_1>0$. Compactness of $I$ thus
imply that hypothesis (i) of Lemma~\ref{lem-criteria} holds. Hypothesis (ii)
follows from the oscillation theorem as stated in Theorem~\ref{theo-osci}. In
fact, as $\mu$ increases each phase $\theta^E_{N,l}(\mu)$ decreases. The
integral in (ii) is the total phase (sum of all $\theta^E_{N,l}(\mu)$'s, in
units of $2\pi$) accumulated as $\mu$ varies in $I$. If $L+K$ eigenvalues pass
by $E$ as $\mu$ varies, the total phase has to change by at least $K$. Hence
the hypothesis of Theorem~\ref{theo-specav} imply (ii) of
Lemma~\ref{lem-criteria}. \hfill $\Box$


\vspace{.2cm}


\section*{Appendix A: reminder on M\"obius transformations}
\label{sec-Moeb}

This appendix resembles the basic properties of the M\"obius transformation as
they are used in the main text. All
proofs are contained in \cite{SB}. Complex matrices of size $2L\times 2L$ are denoted by
mathcal symbols, those of size $L\times L$ by roman letters.

\vspace{.2cm}

The upper half-plane and unit disc (also called Cartan's first classical
domain) are defined by
$$
\UM_L
\,=\,
\left\{
Z\in\mbox{Mat}(L\times L,\CC)
\;\left|\;
\imath(Z^*-Z) >0\;
\right\}
\right.
\,,
\quad
\DM_L
\,=\,
\left\{
U\in\mbox{Mat}(L\times L,\CC)
\;\left|\;U^*U<{\bf 1}
\;
\right\}
\right.
\,,
$$
where $Y>0$ means that $Y$ is positive definite. If $Z\in\UM_L$, then $Z$
is invertible and $-Z^{-1}\in\UM_L$. Moreover, for any $V=V^*$ and any
invertible $T$, one has $Z+V\in\UM_L$ and $T^*ZT\in\UM_L$. The formulas
\begin{equation}
\label{eq-dischalfplane}
U\;=\;(Z\,-\,\imath\,\one)(Z\,+\,\imath\,\one)^{-1}
\;,
\qquad
Z\;=\;\imath\;(\one\,+\,U)(\one\,-\,U)^{-1}
\;,
\end{equation}
establish an analytic diffeomorphism from $\UM_L$ onto $\DM_L$. The
boundary $\partial\DM_L$ of $\DM_L$ is a stratified space given as the
union of strata $\partial_l \DM_L$, $l=1,\ldots,L$, where $\partial_l
\DM_L$  is the set of matrices $U$  for which $U^*U\leq \one$ and
rank$(\one-U^*U)=L-l$. By Proposition~\ref{prop-parametrize} the maximal boundary
$\partial_L\DM_L=\mbox{U}(L)$ is identified with the Lagrangian
Grassmannian $\LM_L$. Similarly, the boundary of $\UM_L$ is stratified,
but this will play no role here.

\vspace{.2cm}

The M\"obius transformation (also called canonical transformation or fractional
transformation) is defined by
\begin{equation}
\label{eq-moebius}
\Tt\cdot Z
\;=\;
(AZ+B)\,(CZ+D)^{-1}
\,,
\qquad
\Tt
\,=\,
\left(
\begin{array}{cc} A & B \\  C & D \end{array}
\right)
\;\in\;\mbox{GL}(2L,\CC)
\,,
\;\;Z\in\mbox{Mat}(L\times L,\CC)\,,
\end{equation}
whenever the appearing inverse exists. For $\Tt$ as in \eqref{eq-moebius} and
as long as the appearing inverse exists, the inverse M\"obius transformation is
defined by
\begin{equation}
\label{eq-moebiusinv} W:\Tt \;=\; (WC-A)^{-1}\,(B-WD) \,, \qquad
W\in\mbox{Mat}(L\times L,\CC)\,.
\end{equation}
The M\"obius transformation is a left action, namely $(\Tt\Tt')\cdot
Z=\Tt\cdot(\Tt'\cdot Z)$ as long as all objects are well-defined.
It is well-known that if $\Tt\in$SP$(2L,\CC)$ and $Z\in\UM_L$, the M\"obius transformation $\Tt\cdot Z$ is well-defined.

\vspace{.2cm}

\section*{Appendix B: Cauchy formula for Cartan's classical domain}
\label{sec-Cauchy}

The results of this section are proven in \cite{Hua}. Let
$\overline{\DM_L}$ be the topological closure of $\DM_L$ and let $dU$
be the normalized Haar measure on its maximal boundary
$\partial_L\DM_L=\,$U$(L)$.

\begin{theo}
For any continuous function $f:\overline{\DM_L}\to\CC$ which is analytic on
$\DM_L$, one has for all $Z\in\DM_L$,
\begin{equation}
\label{eq-Cauchy} f(Z)\;=\; \int dU\; \det(\one-ZU^*)^{-L}\;f(U)\;.
\end{equation}
\end{theo}

For the proof of Theorem~\ref{theo-boundav} we only need the case $Z=\nul$.
Moreover, only intervene functions $f$ which are of the form $f(U)=F(\Cc^*\cdot
U)$ where $\Cc^*\cdot U$ is the Cayley transform of a unitary matrix (written
in the notations of the appendix) and hence hermitian, and $F$ is a complex
function on the hermitian matrices. The Cayley transform $\Cc^*\cdot U$ is not
defined for all unitaries $U$, but it is defined on a set of full measure. The
change of variables formula to the Lebesgue measure $d\xi$ on (real and
imaginary part of each entry of) the Hermitian matrices is now
$$
\int dU\;F(\Cc^*\cdot U)\;=\;c\;\int d\xi\;\det(\one+\xi^*\xi)^{-L}\;F(\xi)\;,
$$
where $c$ is a normalization constant (which is given in \cite{Hua}).
As the
measure $d\xi\,\det(\one+\xi^*\xi)^{-L}$ is invariant under the reflection
$\xi\mapsto -\xi$, it follows that
\begin{equation}
\label{eq-reflext} \int dU\;F(\Cc^*\cdot U)\;=\;\int dU\;F(-\,\Cc^*\cdot U) \;.
\end{equation}

\vspace{.2cm}


\end{document}